# Perfect anomalous reflectors at optical frequencies

Tao He,[1,2,3†] Tong Liu,[4†] Shiyi Xiao,[5] Zeyong Wei,[1,3] Zhanshan Wang,[1,2,3*] Lei Zhou,[4*] Xinbin Cheng,[1,2,3*]

[1] MOE Key Laboratory of Advanced Micro-Structured Materials, Shanghai 200092, China

[2] Institute of Precision Optical Engineering, School of Physics Science and Engineering, Tongji University, Shanghai 200092, China

[3] Shanghai Institute of Intelligent Science and Technology, Tongji University, Shanghai 200092, China

[4] State Key Laboratory of Surface Physics, Key Laboratory of Micro and Nano Photonic Structures (Ministry of Education), and Department of Physics, Fudan University, Shanghai 200438, China

[5] Key Laboratory of Specialty Fiber Optics and Optical Access Networks, Joint International Research Laboratory of Specialty Fiber Optics and Advanced Communication, Shanghai University, Shanghai 200444, China

† T. He and T. Liu equally contributed to this work.
* Corresponding email: wangzs@tongji.edu.cn, phzhou@fudan.edu.cn, chengxb@tongji.edu.cn.

**Abstract**

Reflecting light to a pre-determined non-specular direction is an important ability of metasurfaces, which is the basis for a wide range of applications (e.g., beam steering/splitting and imaging). However, anomalous reflection with 100% efficiency has not been achieved at optical frequencies in conventional metasurfaces, due to losses and/or insufficient nonlocal control of light waves. Here, we propose a new type of all-dielectric quasi-three-dimensional subwavelength structures, consisting of multilayer films and specifically designed meta-gratings, to achieve perfect anomalous reflections at optical frequencies. A complex multiple scattering process was stimulated by effectively coupling different Bloch waves and propagating waves in the proposed meta-system, thus offering the whole meta-system the desired nonlocal control on light waves required to achieve perfect anomalous reflections. Two perfect anomalous reflectors were designed to reflect normally incident 1550 nm light to the 40° and 75° directions with absolute efficiencies higher than 99%, and were subsequently fabricated and experimentally demonstrated to exhibit efficiencies 98% and 88%, respectively. Our results pave the way towards realizing optical meta-devices with desired high efficiencies in realistic applications.

Metasurfaces, ultra-thin metamaterials composed of planar microstructures with tailored optical responses, have gained much attention for their potential for controlling the amplitudes, phases, and polarizations of light waves[1-5]. Many new functions and applications have been demonstrated based on metasurfaces, such as anomalous deflections of light[6-9], dispersion compensation[10, 11], metalens imaging[12, 13], holographic imaging[14, 15], and so forth[16-19]. Recently, optical metasurfaces/meta-devices have attracted increasing attention[20-23], as they are ultra-compact and flat, being ideal replacements for conventional devices in future integration-optics applications. However, the efficiency of metasurfaces is a long-standing bottleneck issue that hinders practical applications of metadevices. For example, although anomalous reflection has been demonstrated as an important capability of metasurfaces, the reported highest efficiency is below 85% at optical frequencies[24], let alone other meta-devices with more sophisticated functionalities. Perfect anomalous reflectors at optical frequencies are extremely desired because they are the basis for highly efficient manipulations of complex optical waves and fields[25-27].

Pioneering works on anomalous deflection have been carried out by engineering the phase gradient along metasurfaces, known as the generalized Snell's law[28, 29]. However, while the reflection-type gradient metasurfaces proposed in Ref. [29] can exhibit very high working efficiencies in the microwave region, the efficiencies of optical metasurfaces thus designed drop significantly, especially for large bending angles. Many studies have focused on how to improve the efficiency of anomalous reflection for large bending angles, from the aspects of mechanism, configuration and optimization[24, 30, 31]. A rigorous analysis by Alu *et al*. suggested that a metasurface should exhibit certain "gain & loss" responses to achieve the desired perfect anomalous reflection[32], which is unfortunately difficult to implement in practice. Other approaches that do not involve "gain & loss" materials were subsequently proposed, which typically take surface waves or evanescent waves into consideration, making the metasurface exhibit the desired effective nonlocal responses yielding similar effects as those of the "gain & loss" metasurfaces[33-35]. However, while perfect anomalous reflectors were realized in the microwave regime using this design scheme[36], direct extensions to optical frequencies are not possible since these metasurfaces involve metallic structures that are highly lossy at optical frequencies[37]. Moreover, although many dielectric metasurfaces have been achieved in the optical domain, they typically suffer from transmission losses and do not necessarily exhibit the desired strong nonlocal responses. Therefore, achieving perfect anomalous reflections at optical frequencies is still a challenging task for nanophotonic community.

Here, we propose an all-dielectric quasi-three-dimensional subwavelength structure (Q3D-SWS), composed of a purposely designed multilayer film and a dielectric meta-grating separated by a dielectric spacer, to achieve the desired perfect anomalous reflection at optical frequencies. These all-dielectric systems are not only free from absorption losses but also eliminate all transmission losses using the multilayer high-reflection films. Moreover, multiple scattering of light inside the Q3D-SWS can be controlled by altering both the reflection phases and propagation phases of light through fine-tuning the multilayer structure and adjusting the spacer thickness. Therefore, optimizing these structural parameters can help us find a perfect-efficiency anomalous reflector, which exhibit the desired nonlocal control ability on light waves assisted by the specifically "designed" multiple-scattering processes. As a proof of concept, we design two perfect-efficiency all-dielectric meta-reflectors working at 1550 nm that can reflect normally incident light to the 40° and 75° directions, and fabricate out the samples and experimentally demonstrate their anomalous reflections with absolute efficiencies 98% and 88%, respectively.

**Models and theory of perfect anomalous reflection**

According to Huygens' principle, far-field outgoing waves are dependent on the secondary sources located in the near field. Therefore, we can tailor the reflection of a metasurface by purposely designing the field distributions (energy-flow distribution, the $z$ component of the Poynting vector) generated on a plane right above the metasurface[31], as shown in Fig. 1(a). A conventional anomalous-reflection metasurface is designed to exhibit a linearly varying reflection-phase distribution. However, since the reflection phase cannot increase to infinity, usually a super-periodicity has to be introduced so that the meta-reflector is actually a grating composed of identical super-cells exhibiting linearly varying phase distribution covering a $2\pi$ range. In general, the system is designed to allow only three reflection channels alive (i.e., the specular reflection and two first-order diffraction channels, see Fig. 1(a)) with all high-order channels carrying evanescent modes. Supposing that $r^{-1}$, $r^0$, $r^{+1}$, $\varphi^{-1}$, $\varphi^0$, and $\varphi^{+1}$ represent the amplitudes and phases of these three reflective channels, we can easily obtain the actual energy-flow distribution by $S_{pz,z=0}(x) = \frac{1}{2}Re(E_{total,x}H^*_{total,y})$ on a plane right above the metasurface,

$$S^a_{pz,z=0}(x) = \frac{1}{2}y_0 E_i^2 \left\{ Re \begin{bmatrix} \left(1-r^0 e^{j\varphi^0}\right)r^{-1}e^{+jGx-j\varphi^{-1}} + \left(1-r^0 e^{j\varphi^0}\right)r^{+1}e^{-jGx-j\varphi^{+1}} \\ -r^{-1}\cos\theta_r e^{-jGx+j\varphi^{-1}}\left(1+r^0 e^{-j\varphi^0}\right) \\ -r^{+1}\cos\theta_r e^{+jGx+j\varphi^{+1}}\left(1+r^0 e^{-j\varphi^0}\right) \\ -2r^{-1}r^{+1}\cos\theta_r \cos\left(2Gx - j\varphi^{-1} + j\varphi^{+1}\right) \end{bmatrix} \right\} \quad (1)$$

which can be further simplified to $S^a_{pz,z=0}(x) = A^a_1 cos(Gx + \tau^a_1) + A^a_2 cos(2Gx + \tau^a_2)$, as shown in Fig. 1(a). Here, $G$ is the reciprocal lattice vector, and $\theta_r$ is the reflection angle of first-order diffraction channels. Here, we have neglected the contributions from the high-order evanescent waves in deriving Eq. (1).

We find that the energy-flow distribution contains two terms. While the first-order oscillation term is originated from the interference between the 0[th] and ±1[st] channels, the second term, however, comes from the interference between -1[st] and +1[st] channels. To realize the desired perfect anomalous reflection as shown in the bottom part of Fig. 1(a), $r^0$ and $r^{-1}$ should be minimized while $r^{+1}$ should take a specific value of $1/\sqrt{cos\theta_r}$ dictated by the law of energy conservation. In other words, the coefficients of actual energy-flow distribution should be $A^a_2 = 0$, $A^a_1 = \frac{1}{2}y_0 E_i^2 \left(\frac{1}{\sqrt{cos\theta_r}} - \sqrt{cos\theta_r}\right)$ and $\tau^a_1 = \varphi^{+1}$, making the actual energy-flow distribution taking the form of

$$S_{pz} = \frac{1}{2}y_0 E_i^2 \left(\frac{1}{\sqrt{cos\theta_r}} - \sqrt{cos\theta_r}\right) cos\left(Gx + \varphi^{+1}\right). \quad (2)$$

Detailed derivations of Eqs. (1)-(2) are given in SI 1.

Equation (2) contains profound physics and useful guidance for us to design appropriate meta-reflections. We note that such a meta-reflector does not conserve energy at a local point *x* (evidenced by the in-balance of input and out-going energy-flow at any local *x* point), but rather conserves the total energy within a whole super-cell since $\int_0^P S_{pz}(x)dx = 0$. Previously, Alu *et al.* proposed that this requirement can be realized in a system composed of gain-lossy materials, which imposed difficulties in realistic device fabrications, especially at optical frequencies. Here, we look for *passive* non-absorbing systems that can satisfy Eq. (2). According to the law of energy conservation, we find that

$$\int_{-z_0}^{0} S_{px}(x_0, z)dz - \int_{-z_0}^{0} S_{px}(x_0 + \Delta x, z)dz = \int_{x_0}^{x_0+\Delta x} S_{pz}(x, z=0)dx \approx S_{pz}(x) \cdot \Delta x \quad (3)$$

where $z=-z_0$ is the bottom plane of system. Equation (3) clearly indicates that the system must exhibit the desired position-dependent lateral energy-transfer capability (i.e., $S_{px}(x) \neq 0$), which also suggests that such a system must possess appropriate non-local responses.

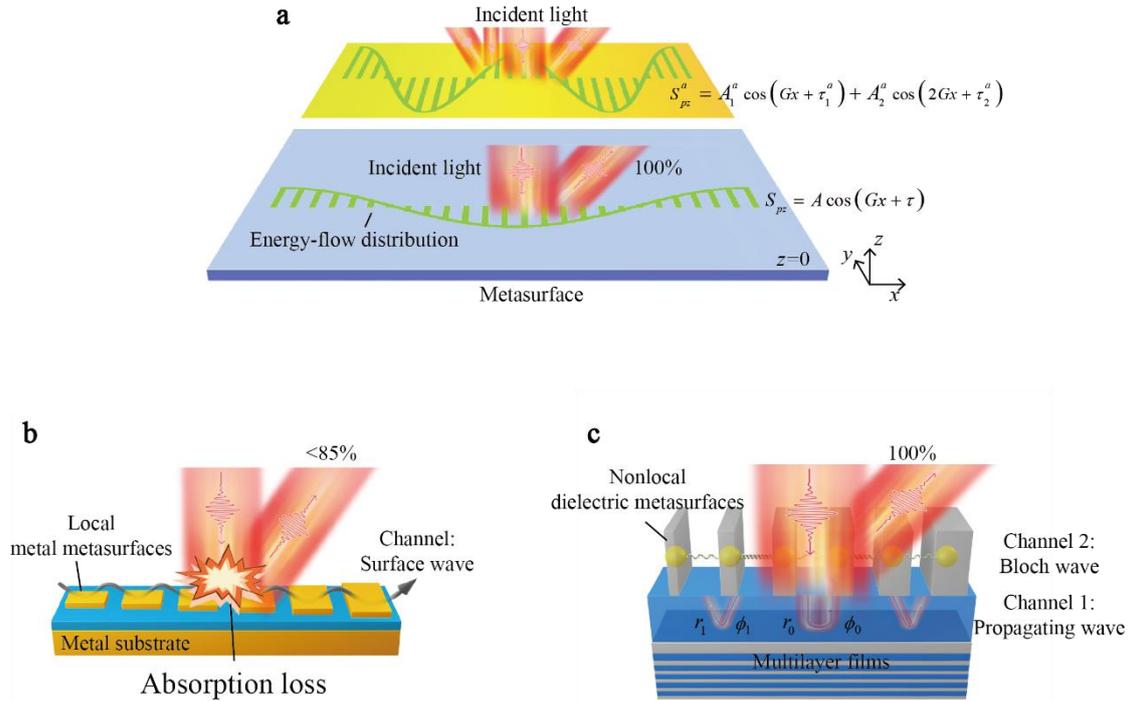

**Figure 1 | Nonlocal response model of meta-reflectors. a**, The energy-flow distribution above the metasurface determines the reflection of light. The energy-flow distribution of imperfect anomalous reflections is composed of both first-order and second-order oscillations, while that of perfect anomalous reflections comprises a first-order oscillation only. The energy-flow distribution above the metasurface is modulated by lateral energy transfer. **b**, Traditional metal metasurfaces, which use local metal metasurfaces and a surface wave to realize the required lateral energy transfer, but the loss of metal in the optical frequencies causes the efficiency to remain below 85%. **c**, Q3D-SWSs use multilayer films to circumvent transmission/absorption loss and control the propagation waves in the spacer and Bloch waves inside nonlocal dielectric metasurfaces to realize perfect anomalous reflectors at optical frequencies. The amplitude and phase responses ($r_0$, $r_1$, $\phi_0$, $\phi_1$) of the multilayer films to the $0^{th}$- and $1^{st}$-order propagation waves can regulate the lateral energy transfer generated by the propagation waves and Bloch waves.

Now the question is: what kind of system exhibits such desired lateral energy-transfer ability? Metasurfaces with the metal-insulator-metal (MIM) configuration have been widely used to regulate the lateral energy transfer of light utilizing the surface wave channel[36, 37], as shown in Fig. 1(b). Perfect anomalous reflections in the microwave region have been demonstrated using this design scheme[36]. However, this design scheme fails to realize perfect anomalous reflections in the optical frequency domain due to the absorption loss of metal, resulting in the reported highest efficiency being below 85%[37].

All-dielectric metasurfaces are free of absorption loss, but they typically suffer from transmission losses and/or are insufficient to provide the desired lateral energy transfer to realize perfect anomalous reflection[38]. Here, combining the advantages of MIM configuration and dielectric metasufaces, we propose a Q3D-SWS which consists of a transmissive dielectric metasurface and a multilayer film separated by a dielectric spacer. The multilayer structure serves as a Bragg mirror well replacing the ground metallic plane in the MIM configuration, while the all-dielectric structures can minimize the absorption loss. Moreover, multiple scattering processes inside the spacer can significantly enhance the lateral energy transfer capability of the whole structure (see Fig. 1(c)). Specifically, as the meta-system is shined by a plane wave, incident light will first be coupled into Bloch waves of different orders inside the top-layer dielectric metasurface, which are then scattered into the spacer as propagating or evanescent waves with tangential wave-vectors matching those of the Bloch waves. It is important to note that, all these waves (e.g., Bloch waves, propagating waves and evanescent waves) contribute to the desired lateral energy transfer. Here, to fully utilize the multiple scatterings to enhance the lateral energy transfer capability, we propose to use a thick spacer (with thickness larger than a quarter-wavelength of light inside the spacer) to construct our meta-system, which is quite different from the usual MIM cases[24, 36]. In such a case, contributions from evanescent waves are relatively small since they cannot be enhanced by multiple scatterings. Therefore, in our Q3D-SWS configuration, the lateral energy transfer $S^a$ is mainly contributed by propagating-wave-induced lateral energy transfer $S^P$ and Bloch-wave-induced lateral energy transfer $S^B$, which can be approximately written as

$$S^a \approx S^P + S^B \qquad (4)$$

The lateral energy transfer is reinforced by the multiple scattering process, which can be described by $\rho_n = R_{n0} + \sum_{m,m'} \widetilde{T}_{nm'} G_{m'm} T_{m0}$, where $R_{n0}$, $\widetilde{T}_{nm'}$ and $T_{m0}$ denote the scattering coefficients of light through the upper-layer metasurface ($R_{n0}$ represents the $n^{th}$ reflection coefficient as a $0^{th}$-order plane wave is incident from air; $\widetilde{T}_{nm'}$ and $T_{m0}$ represent the transmission coefficients as a plane wave is incident from the spacer region or from the air region, respectively) and $G_{m'm}$ are the Green's function elements describing how the $0^{th}$-order incoming wave finally changes to the $n^{th}$-order out-going wave aided by all multiple-scattering processes inside the spacer. $G_{m'm}$ strongly depend on the properties of the multilayer films and the spacer, and are responsible for the complicated multiple scatterings inside the meta-system. A simple expectation is that light wave scattered to the desired direction can be possibly enhanced significantly if an interference can be formed with appropriate phase accumulated for the corresponding optical path.

We now quantify the above physical argument, starting from finding simplified expressions of $S^P$ and $S^B$ exhibiting clear physical meanings. Retaining only the $0^{th}$- and $\pm 1^{st}$-order propagating waves inside the spacer and assuming that the multilayer film can perfectly reflect all these modes (with different reflection phases), we find the following (approximate) formula for $S^P$:

$$S^P = \alpha \operatorname{Re}\left[ \begin{array}{l} \left(1-e^{j\phi_0+j\varphi_{s,0}}\right)\left(1+e^{-j\phi_1-j\varphi_{s,1}}\right)\rho_1(Gx) \\ +\left(1+e^{-j\phi_0-j\varphi_{s,0}}\right)\left(1-e^{j\phi_1+j\varphi_{s,1}}\right)\rho_2(Gx) \end{array} \right] \quad (5)$$

which can be further simplified to a first-order cosine-form lateral energy transfer that has the same form as that of the desired energy-flow distribution, $S^P(x) = A_1^P \cos(Gx + \tau_1^P) + O$, with $O$ representing all high-order contributions. Here, $\alpha$ denotes the amplitude of lateral energy transfer $S^P$ and $\rho$ denotes the abbreviation of oscillation form. Detailed derivation of Eq. (5) can be found in SI 2. Specifically, $\varphi_{s,0}$ and $\varphi_{s,1}$ denote the phase accumulations of $0^{th}$- and $\pm 1^{st}$-order propagating waves in the spacer, while $\phi_0$ and $\phi_1$ are the reflection phases of those two propagating waves on the multilayer film. Clearly, these phases have direct influences on the final $S^P$, which is expected since they can affect the resonance-interference conditions of different optical paths, as shown in Fig. 1(c).

Moreover, $\phi_0$ and $\phi_1$ also indirectly affect the lateral energy transfer contributed by Bloch waves $S^B$, since field distribution inside the spacer can impact the field distribution inside the metasurface according to the boundary conditions. An important observation is that $S^B$ is composed of only first-order and second-order cosine functions, derived as $S^B(x) = A_1^B \cos(Gx + \tau_1^B) + A_2^B \cos(2Gx + \tau_2^B) + O$, because high-order Bloch waves cancel each other and do not contribute to the lateral energy transfer at positions deep inside the dielectric metasurface (details given in SI 2).

We have now established the relationships between four light phases $\phi_0$, $\phi_1$, $\varphi_{s,0}$ and $\varphi_{s,1}$ and two lateral energy transfer functions $S^P$ and $S^B$. We note that the two reflection phases $\phi_0$ and $\phi_1$ are predominantly dictated by the multilayer structure, while the two propagation phases $\varphi_{s,0}$ and $\varphi_{s,1}$ are solely determined by the spacer thickness. Therefore, given a top-layer meta-structure, one can always efficiently control the two lateral transfer functions $S^P$ and $S^B$ via varying the spacer thickness and fine-tuning the multilayer structure. Further, we note that varying the spacer thickness can simultaneously modify $\varphi_{s,0}$ and $\varphi_{s,1}$, being highly undesired for our optimization aiming to enhance the anomalous-reflection channel and suppress the normal-reflection channel. Therefore, here we choose to fix the spacer thickness (and thus fix $\varphi_{s,0}$ and $\varphi_{s,1}$ accordingly) and only fine-tune the multilayer structure to change $\phi_0$ and $\phi_1$ independently in our optimizations, in order to regulate the actual energy-flow distribution $S^a=S^P+S^B$ at will, finally achieving the desired perfect anomalous reflection. Specifically, we optimize the thicknesses of four upper layers of the multilayer film to efficiently change the two reflection phases $\phi_0$ and $\phi_1$, but well maintain the perfect-reflection properties of the whole structure since the remaining part of the multilayer structure is still a well-defined photonic Bragg mirror.

**40° perfect anomalous reflection at 1550 nm**

We first design a perfect anomalous reflector working at 1550 nm wavelength, that can reflect normally-incident TM light to 40° off-normal direction. Considering that the phase requirement of 40° perfect anomalous reflection is still approximately linear, six Si gradient gratings with widths ranging from small to large were first selected. Then, the thickness of the spacer was determined to be 280 nm (other larger values are also feasible); moreover, the effect of evanescent waves was suppressed well, and only propagating waves were allowed to contribute to the lateral energy transfer in the spacer. Multilayer films were then preliminarily selected to totally reflect the $0^{th}$- and $\pm1^{st}$-order propagating waves in the spacer that are scattered by the gradient gratings. Phase responses $\phi_0$ and $\phi_1$ were then scanned to control the oscillations of $S^P$ and $S^B$ of each order and then to regulate the actual energy-flow distribution $S^a$:

$$S^a = S^P + S^B = A_1^a(\phi_0, \phi_1)\cos(Gx + \tau_1^a) + A_2^a(\phi_0, \phi_1)\cos(2Gx + \tau_2^a) \quad (6)$$

The purpose was to minimize the second-order energy-flow oscillation and to match the first-order energy-flow oscillation with the desired form.

According to the diffraction formula, it is easy to determine that the $0^{th}$- and $1^{st}$-order propagating waves in the spacer correspond to incident angles of 0° and 25.9°, respectively. To find a proper $\phi_0$ and $\phi_1$ to realize the desired regulation, the relation of phase responses $\phi_0$ and $\phi_1$ to the second-order amplitude of actual lateral energy transfer $A_2^a$ is exhibited in Fig. 2(a). The phase responses $\phi_0$ and $\phi_1$ corresponding to a low-level amplitude of $A_2^a$ are the target values. It is also worth noting that the amplitude of the second-order oscillation $A_2^a$ is mainly regulated by $\phi_1$, as shown in Fig. 2(a), because $\phi_1$ directly affects the interference of the first-order reflection. The relation of $\phi_0$ and $\phi_1$ with respect to $A_1^a/A$ is exhibited in Fig. 2(b), where $A$ is the desired value of the first-order amplitude. $\phi_0$ and $\phi_1$ have a strong ability to control the amplitude of the first-order oscillation, and the target value of $A_1^a$ is in the green region in Fig. 2(b). It is easy to obtain the proper $\phi_0$ and $\phi_1$ to realize the desired $A_2^a$ and $A_1^a$ simultaneously, as shown by the green pentagram.

Considering that the desired $A_2^a$ and $A_1^a$ are the necessary conditions, Fig. 2(c) shows the anomalous reflection efficiency versus the phase responses $\phi_0$ and $\phi_1$. For the parameters marked by the green pentagram, perfect anomalous reflection is achieved. With a specific combination of $\phi_0$ and $\phi_1$, the anomalous reflection efficiency can be controlled from near 0% to perfect anomalous reflection at will. In addition, the influences of phase responses $\phi_0$ and $\phi_1$ on the parameters $S^P$ and $S^B$ are demonstrated in SI 3. It can be seen from the results that the phase responses $\phi_0$ and $\phi_1$ also have a strong ability to control $S^P$ and $S^B$.

When proper parameters such as (173°, 170°) for $\phi_0$ and $\phi_1$ are selected, the distributions of the second- and first-order oscillations are as shown in Fig. 2(d). The second-order oscillation is reduced to a level that is low enough, and the first-order oscillation of actual energy-flow distribution $S^a$ is close to a perfect energy-flow distribution to realize an anomalous reflection with an efficiency higher than 99%.

Now, the key question is whether it is possible to use multilayer films to realize arbitrary phase combinations of $\phi_0$ and $\phi_1$ while maintaining the high reflectance of the $0^{th}$- and $1^{st}$-order propagating waves. The Bragg mirror in Fig. 2(e) can offer near 100% reflectance only with proper proportions of high- and low-refractive-index materials. The Bragg mirrors that meet the high reflectance requirements fail to realize an arbitrary phase combination because the phase responses $\phi_0$ and $\phi_1$ are associated, as shown in Fig. 2(f). The phase combinations realized by the Bragg mirrors have only a limited scope and are framed in a dotted box. We propose the use of aperiodic multilayer films, as shown in Fig. 2(g), that are composed of a periodic Bragg mirror ($(HL)^7H$) and four phase layers (DCBA) to realize an arbitrary phase combination. The standard Bragg mirror offers nearly 100% reflectance. The phase layers could control $\phi_0$ and $\phi_1$ independently by changing the thickness to regulate the light interference. As shown in Fig. 2(f), the green marks representing the phase combination realized by changing the thickness of the four phase layers occupy the whole phase space. Therefore, the arbitrary phase combination ($\phi_0$, $\phi_1$) and near-100% reflectance can be obtained simultaneously. These aperiodic multilayer films greatly expand the ability of Q3D-SWSs.

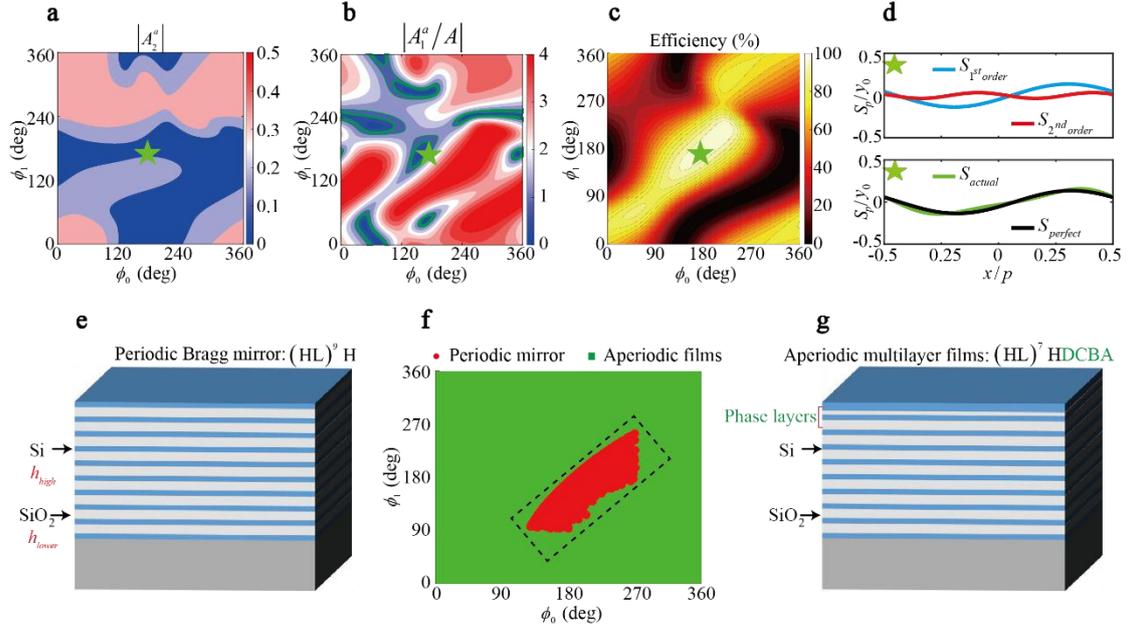

**Figure 2 | Nonlocal design for 40° perfect anomalous reflection.** The nonlocal response of perfect anomalous reflection is realized by changing only the phase responses $\phi_0$ and $\phi_1$ of multilayer films with a given spacer and gradient gratings. The geometric parameters of the six elements are listed. The height of the six elements is 500 nm, and the widths are 71 nm, 161 nm, 201 nm, 217 nm, 236 nm, and 356 nm. **a**, The relation of phase responses $\phi_0$ and $\phi_1$ with respect to the second-order amplitude of actual lateral energy transfer $S^a$. **b**, The relation of phase responses $\phi_0$ and $\phi_1$ with respect to the first-order amplitude of actual lateral energy transfer $S^a$. The actual first-order amplitude is normalized by the desired amplitude. **c**, The anomalous reflection efficiency versus phase responses $\phi_0$ and $\phi_1$. The perfect anomalous reflection is marked by the green pentagram. **d**, Distributions of the second- and first-order oscillations when $\phi_0$ and $\phi_1$ are (173°, 170°), respectively. **e**, Schematic diagram of the periodic Bragg mirror. **f**, Phase combination realized by the periodic Bragg mirror and aperiodic multilayer films. The phase responses $\phi_0$ and $\phi_1$ of the Bragg mirror are change with the thickness. For aperiodic multilayer films, an arbitrary phase combination ($\phi_0$, $\phi_1$) and near-100% reflectance can be obtained simultaneously by changing the thickness of the four outermost phase layers. **g**, Schematic diagram of the aperiodic multilayer films. The periodic Bragg mirror and four phase layers control the amplitude and phase, respectively.

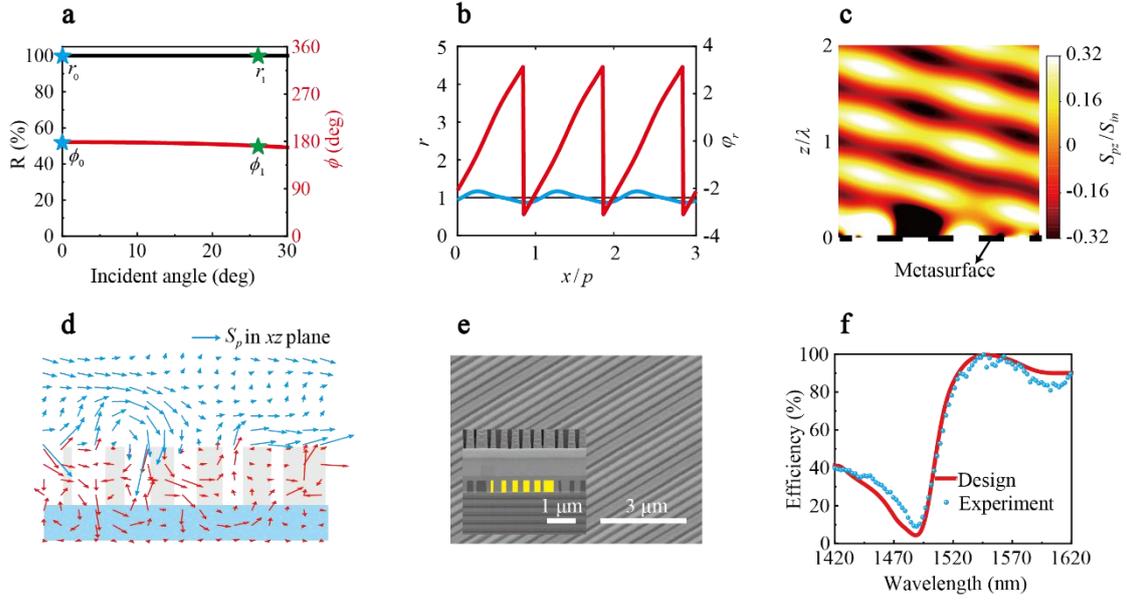

**Figure 3 | Analysis and preparation for 40° perfect anomalous reflection**. **a,** Amplitude and phase responses of the designed multilayer films for different incident angles. **b,** Actual reflection coefficients of the whole meta-system, including the amplitude $r(x)$ and reflection phase $\varphi_r(x)$. **c,** Actual energy-flow distribution of $S_{pz}$. An interference pattern is produced by the incident wave and perfect anomalous reflection wave. **d**, The arrows represent the direction of $S_p$ in the *xz*-plane. Lateral energy transfers exist in both the spacer and metasurfaces. **e,** SEM side-view image and cross-section image of the sample. **f,** Broadband spectra measured in the experiment and calculated in the design.

Finally, a proper multilayer stack was designed to realize the corresponding $\phi_0$, $\phi_1$ and high reflectance simultaneously. The designed films have near-100% reflectance at 0° and 25.9° incident angles. At the same time, the responses of the phase to the 0$^{\text{th}}$- and 1$^{\text{st}}$-order propagation waves are near (173°, 170°), respectively. Fig. 3(a) shows the amplitude and phase response of the films. The actual reflection coefficients of the whole meta-system, including the amplitude $r(x)$ and reflection phase $\varphi_r(x)$, are exhibited in Fig. 3(b). The amplitude is greater than 1 in some regions and less than 1 in other regions corresponding to the required nonlocality, and the phase is nearly linear, confirming the previous assumptions. To further confirm the performance of the Q3D-SWS, the distribution and directions of the Poynting vectors are exhibited in Fig. 3(c) and (d). An expected interference pattern is produced in Fig. 3(c) by the incident wave and perfect (>99%) anomalous reflection wave. As shown by the red arrows in Fig. 3(d), lateral energy transfers exist in both the spacer and metasurfaces. In addition, the blue arrows at the air side are upward in some regions and downward in other regions, corresponding to the nonlocal response "gain & loss". Moreover, because multilayer films have a very strong ability to regulate the amplitude and phase of electromagnetic waves, various film systems satisfy the requirements. The different multilayer films are exhibited in SI 4.

A multilayer film deposition process and microstructure preparation technology were used to prepare the sample. The detailed preparation process is given in the sample fabrication section. Fig. 3(e) shows the side-view image of the sample, and the illustration shows the SEM cross-section image of the sample. The parameters of the prepared sample are consistent with the design. We conducted spectral tests and found that the reflected light at other orders was well suppressed at the designed wavelength of 1550 nm. The detailed spectral tests are discussed in the sample characterization section. Fig. 3(f) shows the anomalous reflection efficiency in the broadband region obtained in the experimental test, up to 98%. For comparison, the designed efficiency in the broadband region is also exhibited in Fig. 3(f). Good agreement indicates the reliability of our fabrication and test results.

**75° perfect anomalous reflection at 1550 nm**

Perfect anomalous reflection for large bending angles is also desired in some important applications, such as large-numerical-aperture metalenses[39, 40] and spectrographs[9]. Therefore, a 75° perfect anomalous reflection of a 1550 nm TM wave with normal incidence was also designed using the Q3D-SWS. As mentioned in SI 5 and reference[32], the phase distribution deviates from the linear gradient, and the requirement of nonlocality increases with the bending angle. Hence, using a gradient grating is not the best choice to realize perfect anomalous reflection with a large bending angle, as shown in SI 5. We proposed the simultaneous selection of $\phi_0$, $\phi_1$ and the width of the sub-unit grating (such as the red grating in Fig. 4(a); the result is similar for other gratings) for scanning to control each order oscillation of $S^P$ and $S^B$ and then regulation of the actual energy-flow distribution $S^a$. $\phi_0$ and $\phi_1$ were realized by designing proper multilayer films.

In the actual design process, considering the fabrication constraints, four unit elements with a period of 400 nm were adopted, and the thickness of the spacer was set as 372 nm. We show the anomalous reflection efficiency versus the three parameters under the condition of 100% reflectance in Fig. 4(b). The anomalous reflection efficiency can be controlled over a wide range at will. The parameters for perfect anomalous reflection (>99%) are indicated by the pentagram in the figure. The exact parameters for $d$, $\phi_0$, and $\phi_1$ are 64 nm, 170°, and 149°. Then, the second- and first-order oscillations of the actual power-flow distribution $S^a$ are shown in Fig. 4(c). The second-order oscillation is reduced to a very low level, and the first-order oscillation is close to a perfect energy-flow distribution. By regulation of the lateral energy transfer in the entire structure, perfect anomalous reflection for large bending angles is realized.

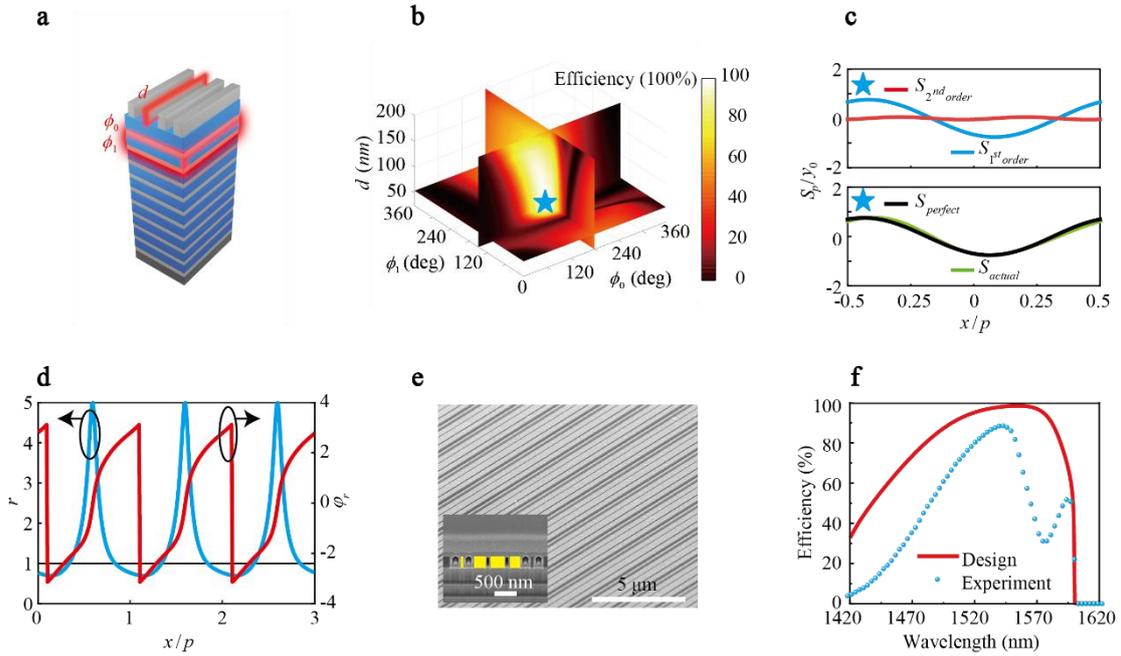

**Figure 4 | Design and analysis of 75° perfect anomalous reflection. a,** The width of the second grating, $\phi_0$, and $\phi_1$ are selected for scanning to realize perfect anomalous reflection at 75°. The geometric parameters of the final four elements are listed. The height of the six elements is 315 nm, and the widths are 174 nm, 64 nm, 264 nm, and 370 nm. **b,** Anomalous reflection efficiency versus the three parameters. The parameters for perfect anomalous reflection are marked by a blue pentagram. The exact parameters for $d$, $\phi_0$, and $\phi_1$ are 64 nm, 170°, and 149°. **c,** Comparison of the actual energy-flow distribution and distribution corresponding to perfect anomalous reflection at 75°. **d,** Actual reflection coefficients of the whole meta-system, including the amplitude $r(x)$ and reflection phase $\varphi_r(x)$. **e,** SEM side-view image and cross-section image of the sample. The holes in the SEM image come from the

insufficient deposition in the FIB process. **f,** Broadband efficiency measured in the experiment and calculated in the design.

Another proper multilayer stack was designed to meet the 100% reflectance and (170°, 149°) phase requirements by changing the thickness of the four outermost phase layers. The detailed parameters and spectra are shown in SI 4. The actual reflection coefficients of the whole meta-system, including the amplitude $r(x)$ and reflection phase $\varphi_r(x)$, are exhibited in Fig. 4(d) to check the performance of the Q3D-SWS. The amplitude is obviously greater than 1 in some regions and less than 1 in other regions, and the phase deviates from a linear gradient. The reflection coefficients are in agreement with the nonlocal requirements of perfect anomalous reflection.

Finally, the same process was used to fabricate the sample. Fig. 4(e) shows the SEM side-view image and cross-section image of the sample. Fig. 4(f) demonstrates the broadband anomalous reflection efficiency in the experiment with a maximum efficiency of 88% and in the design with an efficiency greater than 99%. We speculated that the aforementioned efficiency degeneration comes from the test error for a large reflection angle in the spectral test and from fabrication errors. The efficiency could be further increased to the designed efficiency by optimizing the spectral test and preparation process.

**Discussion and Conclusions**

The reflex response of the meta-system can be designed by controlling the lateral energy transfer, which comes from the propagation wave and Bloch wave in our proposed Q3D-SWS. We mainly discussed how to regulate the lateral energy transfer by sweeping the phase responses $\phi_0$ and $\phi_1$ realized by multilayer films. This proposed strategy is straightforward, non-optimized and effective. Furthermore, the parameters of metasurfaces can help to control the lateral energy transfers of propagation waves and Bloch waves. In a joint design process, a more complicated energy-flow distribution and functionality can be realized by simultaneously optimizing the phase responses and parameters of metasurfaces. For example, we demonstrated a polarization-independent perfect anomalous reflection and a polarization-dependent mirror in Fig. 5(a) and (b) by using a two-dimensional structure (detailed parameters are given in SI 6). The corresponding spectra are shown in Fig. 5(c) and (d). The methodology used to regulate the lateral energy transfer in the Q3D-SWS structure via phase responses $\phi_0$ and $\phi_1$ was applied not only to perfect anomalous reflection but also to other high-efficiency metasurfaces, such as anomalous refractors and metalenses.

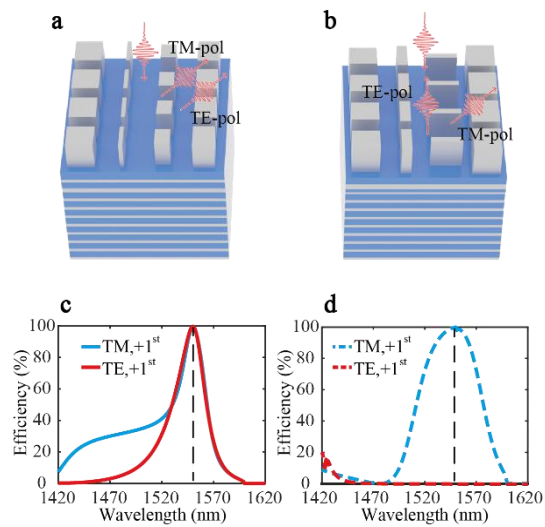

**Figure 5 | Design of complicated functionality via optimization.** Schematic diagram of **a,** polarization-independent perfect anomalous reflection and **b,** a polarization-dependent mirror. The corresponding spectra of the **c,** polarization-independent perfect anomalous reflection and **d,** polarization-dependent mirror. The specular reflection of the structure in figure b is 100% for TE polarization.

In conclusion, we realize a perfect anomalous reflector at optical frequencies for the first time via a new type of Q3D-SWS structure consisting of multilayer films and a specifically designed meta-grating. The multiple scattering process in the Q3D-SWS can be stimulated by adjusting the phase responses of multilayer films, which enhances the lateral energy-flow transfer in the proposed meta-system, finally offering the whole system appropriate nonlocal control of light waves to realize perfect anomalous reflections. The arbitrary phase combination and near-100% reflectance of multilayer films are realized by using aperiodic multilayer films composed of a periodic Bragg mirror and four phase layers. We demonstrate 40° and 75° perfect anomalous reflectors at 1550 nm using our design methodology and structure both in design and in an experiment. The highest anomalous reflection efficiency in both the design and experiment to date (SI 7 shows a comparison of our work with published articles) demonstrates the validity of our proposed methodology and structure. Our work paves the way towards the practical application of metasurfaces in, e.g., spectrograph, laser radar and ranging devices.

**Methods**

1. **Numerical simulation**

   The CST MICROWAVE STUDIO software was used to perform the full-wave simulations. The scattering coefficients of the superstructure and anomalous reflection efficiency were calculated by the S parameters of the CST software. In the simulation, the simulation domain is 2400 nm × 20 nm × 5 μm (along the *x*, *y*, and *z*-direction, respectively), and it corresponds to one period of metasurfaces. The boundary conditions are unit cell at *x* and *y* direction and open at *z*-direction. The refractive indices of Si, $SiO_2$, and substrate are 3.5, 1.48 and 1.51. The system was illuminated by a TM-polarized plane wave. The multilayer film design is carried out by OptiLayer software.

2. **Sample fabrication**

   The glass substrate is first cleaned by ultrasonic cleaning. The $Si/SiO_2$ multilayer films, $SiO_2$ spacer, and Si film are coated using a magnetron sputtering deposition system NSC-15 from Optorun. The power was set as 10 kW and kept constant in the deposition process. The chamber was pumped down to a base pressure of $1.0\times10^{-3}$ Pa and the work pressure was $6.0\times10^{-1}$ Pa. The deposition temperature was room temperature. The current and voltage were automatically regulated. In the process, Si was the only starting material and injected $O_2$ to generate $SiO_2$. The deposition rates under the above condition were 0.25 nm/s and 0.63 nm/s for Si and $SiO_2$, respectively. After deposition, we used TEM and spectral inversion technology to confirm the thicknesses of the multilayer films, and carried out the correction on preparation process and metasurface parameters. Then, the substrate is spin-coated with a positive electron-beam resist (AR6200.09) at a speed of 4000 r.p.m. for 40 seconds and baked at 150 °C on the hotplate for 2 minutes. The final thickness is about 200 nm. The sample is then exposed through electron-beam lithography with 100-kV acceleration voltage. The exposed resist is developed in standard developing liquid. Reactive ion etching (RIE) is performed using $C_4F_8$ and $SF_6$ at 45 and 20 sccm, respectively, at a pressure of 10 mT. RIE power was set to 13 W and Inductively Coupled Plasma power to 400 W. The resist mask was removed by the oxygen plasma.

3. **Sample characterization**

The angle-resolved spectrum system in micro-region (ARM) from Ideaoptics Inc.[41] is employed to verify the spectral property of the 40° anomalous reflection sample. The measurement realizes angle resolution by focusing the reflective beam in the Fourier plane of the lens. The measurement system is not suitable for a 75°-anomalous reflection sample because the measurement system equipped with a microscope objective whose numerical aperture is 0.87 only guarantees a collection angle of 60°. Therefore, the R1, angle-resolved spectrum system, ideaoptics, China is employed to characterize the 75° anomalous reflection sample. The measurement realizes angle resolution by mechanical rotating of the receiver. In the test, the intensity of incident light was detected via the sample with high-reflective films. Then the intensities of +1st, −1st, and 0th order light were detected for all wavelength at once to obtain the anomalous reflection efficiency. The measurements were carried in a darkroom to improve the signal-to-noise ratio.

**Acknowledgements**

This work was supported by the National Natural Science Foundation of China (61925504, 61621001, 62020106009, 62111530053), the National Key Research and Development Program of China (2016YFA0200900), the Major projects of Science and Technology Commission of Shanghai (17JC1400800), the "Shu Guang" project supported by Shanghai Municipal Education Commission and Shanghai Education (17SG22), and the Natural Science Foundation of Shanghai (No. 20JC1414604).

**Author contributions**

X.B.C., L.Z. and Z.S.W conceived and supervised the project. T.H. conducted the numerical simulations with assistance from T.L. and Z.Y.W. on electromagnetic field simulation. T.H. prepared the devices, performed the experiments and analysed the data. T.H., T.L., S.Y.X., Z.S.W, X.B.C and L.Z. conducted the modelling and theoretical analysis. All authors discussed the manuscript. Finally, T.H., T.L., L.Z. and X.B.C wrote the paper with contributions from all authors.

**Competing interests**

The authors declare no competing interests.